\documentclass[aps,prd,eqsecnum,onecolumn,showpacs,nofootinbib,preprintnumbers,superscriptaddress]{revtex4}
\usepackage[T1]{fontenc}
\usepackage[latin9]{inputenc}
\usepackage[letterpaper]{geometry}
\usepackage{bm}
\usepackage{amsmath}
\usepackage{esint}
\usepackage{color}
\usepackage{graphicx}
\PassOptionsToPackage{normalem}{ulem}
\usepackage{ulem}

\makeatletter

\usepackage{times}
\usepackage{hyperref}

\def\beq{\begin{equation}}
\def\eeq{\end{equation}}

\def\m{\hat m}
\def\R{{\cal R}}

\makeatother

\begin{document}

\title{Oscillatory features in the curvature power spectrum after a sudden turn of the inflationary trajectory}

\author{Xian Gao}%
    \email[Email: ]{xgao@apc.univ-paris7.fr}
    \affiliation{%
        \href{http://www.apc.univ-paris7.fr/APC_CS/en}{Astroparticule et Cosmologie (APC)}, UMR 7164-CNRS, Universit\'{e} Denis Diderot-Paris 7,
        10 rue Alice Domon et L\'{e}onie Duquet, 75205 Paris, France}
    \affiliation{%
        \href{http://www.lpt.ens.fr/?lang=en}{Laboratoire de Physique Th\'{e}orique, \'{E}cole Normale Sup\'{e}rieure}, 24 rue Lhomond, 75231 Paris, France,
        }%
    \affiliation{%
        \href{http://www.iap.fr/english/}{Institut d'Astrophysique de Paris (IAP)}, UMR 7095-CNRS, Universit\'{e} Pierre et Marie Curie-Paris 6, 98bis Boulevard Arago, 75014 Paris, France.
        }%

\author{David Langlois}%
    \email[Email: ]{langlois@apc.univ-paris7.fr}
    \affiliation{%
        \href{http://www.apc.univ-paris7.fr/APC_CS/en}{Astroparticule et Cosmologie (APC)}, UMR 7164-CNRS, Universit\'{e} Denis Diderot-Paris 7,
        10 rue Alice Domon et L\'{e}onie Duquet, 75205 Paris, France}
        
\author{Shuntaro Mizuno}%
    \email[Email: ]{shuntaro.mizuno@apc.univ-paris7.fr}
     \affiliation{%
        \href{http://www.apc.univ-paris7.fr/APC_CS/en}{Astroparticule et Cosmologie (APC)}, UMR 7164-CNRS, Universit\'{e} Denis Diderot-Paris 7,
        10 rue Alice Domon et L\'{e}onie Duquet, 75205 Paris, France}
    \affiliation{%
        \href{http://www.th.u-psud.fr}{Laboratoire de Physique Th\'{e}orique, Universit\'{e} Paris-Sud 11 et CNRS}, B\^{a}timent 210, 91405 Orsay Cedex, France}

\date{\today}

\begin{abstract}
In the context  of  two-field inflation characterized by a light direction and a heavy direction, we revisit the question of the impact of the massive modes  on the power spectrum produced after a turn in the inflationary trajectory.  We consider in particular the resonant effect due to the  background oscillations following a sharp turn. Working in the mass basis, i.e. in the basis spanned by the eigenvectors of the effective mass matrix for the perturbations, we provide an analytical estimate of the resonant effect, using  the in-in formalism. In comparison with earlier estimates, we find the same the spectral dependence  but  a smaller amplitude. We also compute, again via the in-in formalism, the effect of the direct coupling between the light and heavy modes at the instant of the turn and confirm our previous results obtained via a different method.  
\end{abstract}

\maketitle

\section{Introduction}
The possibility that high energy effects, with typical scale higher than the Hubble scale during inflation, might leave some imprint in the primordial cosmological perturbations has been intensively explored during the last few years. In particular, it has been realized in the context of multi-field inflation that very massive directions, which were usually thought to be irrelevant during inflation, could have some impact on the cosmological perturbations generated during inflation when the inflationary trajectory in the multi-dimensional field  space is non trivial. Considering the simplest case of a two-dimensional field space with a light direction and a massive direction, it was shown that, for a moderate bending of the trajectory, one can describe the system as a single field model with a reduced effective speed of sound~\cite{Tolley:2009fg}:
\beq
c_s^{-2}=1+4\dot\theta^2/m_{\rm eff}^2\,,
\eeq
where $\theta$ denotes the angle of the inflationary direction in field space (with respect to some fixed basis) and $m_{\rm eff}$ corresponds to the heavy mass plus some corrections ($m_{\rm eff}^2\simeq m_h^2-\dot\theta^2$). The variation of $c_s$ along the inflationary trajectory leads to specific features in the power spectrum~\cite{Achucarro:2010jv,Achucarro:2010da} (see also 
\cite{Gao:2009qy,Chen:2009we,Chen:2009zp,Jackson:2010cw,Cremonini:2010ua,Jackson:2011qg,Behbahani:2011it,Achucarro:2012sm,Jackson:2012fu,Avgoustidis:2012yc,Chen:2012ge,Pi:2012gf,Achucarro:2012yr,Burgess:2012dz,Gwyn:2012mw,Noumi:2012vr,Achucarro:2012fd,Battefeld:2013xka}
for other works investigating the effects of heavy modes from the view point
of a single field effective theory). 
 However, if  sharp  turns occur along  the trajectory, the massive degrees of freedom can be excited, which implies a breakdown of the effective single field description~\cite{Shiu:2011qw,Cespedes:2012hu}.
 A related consequence of a sharp turn is the existence, just after the turn, of oscillations along the massive direction. These oscillations affect the background evolution via a small oscillatory contribution, which can lead to  resonant effects on the  power spectrum and bispectrum of the cosmological perturbations~\cite{Chen:2008wn,Chen:2011zf,Chen:2011tu}, with the potential perspective to detect this signature in the observations~\cite{Chen:2012ja,Jackson:2013mka,Ade:2013uln}.

Note that although we consider here only one heavy degree of freedom in addition to a single light degree of freedom, the case of several heavy degrees of freedom have also been considered \cite{Peterson:2011yt,GLM,Cespedes:2013rda}.

In the present work, we revisit the question of the impact on the curvature power spectrum of the sharp turn and of the subsequent oscillations along the inflationary trajectory. In contrast with most previous analytical treatments, our results take into account both the light and heavy perturbations, without resorting to a single-field approximation. Our treatment relies on the use of the mass basis defined by the eigenvectors of the effective mass matrix (which almost coincides with the Hessian of the potential) instead of the often used adiabatic/entropic decomposition.

 \begin{figure}[h]
\begin{minipage}{0.45\textwidth}
    \centering
        \includegraphics[width=\textwidth]{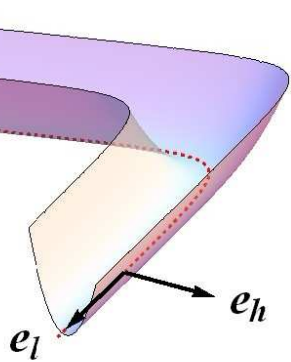}
            \caption{
Schematic illustration of a  valley in some two-dimensional potential. The red dashed line indicates the bottom of the potential. The light and heavy directions associated with the valley are also shown.
}
            \label{fig:valley}
\end{minipage}
\end{figure}

\begin{figure}[h]
\begin{minipage}{0.45\textwidth}
    \centering
        \includegraphics[width=\textwidth]{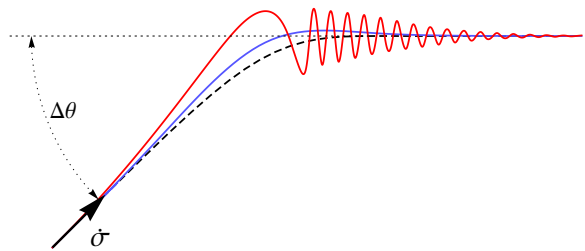}
            \caption{Schematic representation of the inflationary trajectory after a sharp turn (red) or a soft turn (blue). The dashed curve corresponds to the bottom of the valley. 
}
            \label{fig:trajectory}
\end{minipage}
\end{figure}

In the following, we consider a two-field model described by the action
  \begin{equation}
    \label{action}
        S=\int d^{4}x\sqrt{-g}\left(-\frac{1}{2}\delta_{IJ}\partial_{\mu}\phi^{I}\partial^{\mu}\phi^{J}-V\left(\phi^I\right)\right),\qquad \phi^I=\{\phi,\chi\}
    \end{equation}
where $g$ is the determinant of the spacetime metric $g_{\mu\nu}$ and $V(\phi,\chi)$ is the potential of the scalar fields. For simplicity, we have chosen a flat metric in field space, i.e. $G_{IJ}=\delta_{IJ}$ although our treatment could be extended to non trivial field space metrics (see e.g. \cite{LRPST}). 
We suppose that the potential $V(\phi, \chi)$ contains a valley, which suddenly bends along the inflationary trajectory (see Fig.\ref{fig:valley}).
 If the velocity of the inflaton when it reaches the turn is sufficiently large, or equivalently if the turn is sufficiently sharp, then the trajectory will deviate from the local minimum of the potential. After the turn, the inflationary motion will undergo Hubble damped oscillations about the bottom of the valley (see Fig.\ref{fig:trajectory}). 

The goal of the present work is to study in detail, and as much as possible analytically, the effect of this sudden turn on the final curvature power spectrum. One can distinguish two effects that will affect the final curvature power spectrum. The first effect is the direct coupling of the massive modes to the light modes at the turn, which depends on the angular velocity at the turn. The second effect is a resonant amplification of the perturbations arising from the modulation of the background evolution, due to the oscillations about the bottom of the valley after the turn. We have already treated the first effect in a previous work \cite{GLM}, using directly the equations of motions for the perturbations. In the present work, we compute again this effect by resorting to  the in-in formalism, which has now been widely  used in the literature, and thus confirm our previous results. The main emphasis of this paper, however, is the study of the resonant effect. In particular, we show that a full treatment, including both light and massive perturbations, is greatly simplified by using the mass basis. Working in this basis, we compute the correction to the power spectrum due to this effect. In comparison with previous estimates, we find the same spectral dependence  but an amplitude which is much smaller. 

The plan of our paper is the following. In the next section, we summarize the general formalism to treat the background evolution and the linear perturbations in multi-field inflation. We also present a few basic formulas in the context of the in-in formalism  that will be useful for our calculations. The subsequent section, Section \ref{section3}, is devoted to the impact of the background oscillations on the power spectrum. Section \ref{section4} then deals with the effect of the coupling between massive and light modes at the turn. We summarize our results in the conclusion. We have also added an appendix that tries to  compare the various notations, and approximations, used in the literature.

\section{General formalism}
\label{section2}
In this section, we present our general formalism to study the equations of motion for the background and for the linear perturbations in the context of multi-field inflation (see e.g. \cite{Langlois:2010xc} for a pedagogical introduction). In particular, we emphasize the advantage of the mass/potential basis, as introduced in our previous work \cite{GLM}, with respect to the more traditional kinematic basis. We conclude this section by a short introduction to the in-in formalism, which will be used in the following sections.

\subsection{Background equations of motion}
In a spatially homogeneous and isotropic spacetime, endowed  with the  metric
\beq
ds^2=g_{\mu\nu}dx^\mu dx^\nu=-dt^2+a^2(t)\delta_{ij}dx^idx^j,
\eeq
the evolution of  the scale factor $a(t)$ is governed by  the Friedmann equations
    \begin{equation}
H^{2}=\frac{1}{3}\left(\frac{1}{2}\delta_{IJ}\dot{\phi}^{I}\dot{\phi}^{J}+V\right),\qquad\qquad \dot{H}=-\frac{1}{2}\delta_{IJ}\dot{\phi}^{I}\dot{\phi}^{J}\equiv-H^{2}\epsilon,\label{eom_bg}
\end{equation}
where $H\equiv \dot a/a$ is the Hubble parameter and a dot denotes a derivative with respect to the cosmic time $t$.
We work in units such that $M_P\equiv (8\pi G)^{-1/2}=1$ and we adopt Einstein's convention in which repeated indices implicitly denote a summation. The
equations of motion for the homogeneous scalar fields are
\begin{equation}
\ddot{\phi}^{I}+3H\dot{\phi}^{I}+\delta^{IJ}V_{,J}=0\,,\label{phi_eom_pb}
\end{equation}
where $V_{,J}\equiv \partial V/\partial \phi^J$.

As we will see later, it can be convenient to make a change of orthonormal basis in field space, so that the initial velocity components, for instance, are reexpressed as  
    \begin{equation}
        \dot\phi^I = \dot\phi^m \, e^I_m,\qquad \qquad m=1,\cdots,N, \label{change_basis}
    \end{equation}
    where the new basis vectors $e^I_m$ satisfy  the orthonormality condition
    \beq
    \label{orthonormality}
    \delta_{IJ}\, e^I_me^J_n=\delta_{mn},
    \eeq
    as well as  $\delta^{mn}e^I_me^J_n = \delta^{IJ}$.

In the new basis, 
the equations of motion (\ref{phi_eom_pb})  become
\begin{equation}
\label{eom_bck_gen}
         \ddot{\phi}_m +Z_{mn}\dot{\phi}_n+3H \dot{\phi}_m +V_{,m}=0,
    \end{equation}
where  
\beq
\label{Z_mn_def}
 Z_{mn}:= \delta_{IJ} \, e^I_m \dot{e}^J_n\,, \qquad V_{,m}\equiv e^I_m\,  V_{,I}\,.
\eeq
The coefficients $Z_{mn}$, which  are antisymmetric ($Z_{mn}=-Z_{nm}$)   as a consequence of the orthonormality condition (\ref{orthonormality}),  characterize the rotation rate of the new basis with respect to the former field space basis. In a two-dimensional field space, the new basis is fully characterized by its angle $\theta$ with respect to the initial basis, and the coefficients $Z_{mn}$ are simply given by
   \begin{equation}
\bm{Z}=\left(\begin{array}{cc}
0 & -\dot\theta\\
\dot\theta & 0
\end{array}\right).
\label{Z_matrix}
\end{equation}

\subsection{Equations of motion for the linear perturbations}

Let us now turn to the linear perturbations of the two scalar fields coupled to gravity.
For  multi-field models, it is convenient to work in the spatially-flat gauge, where the scalar-type perturbative degrees of freedom are encoded in the scalar field perturbations $Q^I = \delta\phi^I$ (see Appendix for details). The quadratic action for the $Q^I$ can be obtained by expanding the action $S$, given in (\ref{action}), to quadratic order in the perturbations. Using the conformal time $\eta=\int dt/a(t)$ instead of the cosmic time $t$,  the action for the linear perturbations,   expressed in terms of the canonically-normalized variables $v^I=aQ^I$, is given by 
    \begin{equation}
S^{(2)}=\frac{1}{2}\int d\eta \, d^{3}x\left[v'^{T}v'-\delta^{ij}\partial_{i}v^{T}\partial_{j}v-a^{2}v^{T}\left(\bm{M}-H^{2}\left(2-\epsilon\right)\right)v\right],\label{S_canon}
\end{equation}
where we use a matrix notation, $v$ being the column vector with components $v^I$ and $v^T$ is the corresponding transposed matrix. A prime denotes a derivative with respect to the conformal time $\eta$. The matrix  $\bm{M}$ corresponds to  the (squared) mass matrix and is given explicitly by
    \begin{equation}
        M_{IJ}\equiv V_{,IJ}+\left(3-\epsilon\right)\dot{\phi}_{I}\dot{\phi}_{J}+\frac{1}{H}\left(V_{,I}\dot{\phi}_{J}+\dot{\phi}_{I}V_{,J}\right)\,, 
  \label{M_IJ_ori}
\end{equation}
where $\epsilon$ is the standard slow-roll parameter, introduced in (\ref{eom_bg}).

If we go to  another orthonormal basis $e^I_m$, as defined in (\ref{change_basis}),  the linear perturbations will be described by  new components $u^m$, which are related to the former components as 
    \begin{equation}
        v^I = e^I_m u^m,\qquad \qquad m=1, 2\,.\label{modes_trans}
    \end{equation}
    The equation of motion for the $u^m$ can be derived from  the quadratic action
    \begin{equation}
        S=\frac{1}{2}\int d\eta\,  d^{3}x\left[u'^{T}u' + u^{T}\partial^2u-a^{2}u^{T}\left(\tilde{\bm{M}} -H^{2}\left(2-\epsilon\right)\right)u+2au'^{T}\bm{Z}u\right],
        \label{S_mass}
        \end{equation}
with
    \begin{equation}
        \tilde{\bm{M}} := \bm{M} + \bm{Z}^2,
    \end{equation}
where
$\bm{M}$ now denotes the matrix of coefficients  $M_{mn}$ (we use the same notation $\bm{M}$ for simplicity) defined by
    \begin{equation}
        M_{mn}:=M_{IJ}e^I_me^J_n ,
    \end{equation}
and where $\bm{Z}$ is the antisymmetric matrix of components $ Z_{mn}$ introduced in ({\ref{Z_mn_def}).
In matrix notation, these equations of motion are given by
    \begin{equation}
        u''-\partial^2u+a^2\left(\tilde{\bm{M}}-H^2(2-\epsilon) +\dot{\bm{Z}}+H\bm{Z}\right)u+2a\bm{Z}u'=0.\label{eom_full}
    \end{equation}
    The only difference with the equations of motion for the quantities $v^I$ is the presence of the rotation matrix $Z$. 

\subsection{Kinematic basis versus potential basis}
We now present two possible choices for the orthonormal basis, in addition to the original basis defined directly with respect to the scalar fields $\phi^I$. 
\subsubsection{Kinematic basis}
A natural choice of basis is the kinematic basis, corresponding to the usual adiabatic-entropic decomposition \cite{Gordon:2000hv,GNvT}. The adiabatic  direction is represented by the unit vector
\beq
\label{adiab_vector}
n^I= \frac{\dot\phi^I}{\dot\sigma}, \qquad \dot\sigma\equiv \sqrt{\delta_{IJ}\, \dot\phi^I\dot\phi^J},
\eeq
while the entropic direction is along the unit vector $s^I$, which is orthogonal to $n^I$. We call the basis $\{n^I, s^I\}$ the ``kinematic basis'', since $n^I$ is always pointing in the velocity direction ($n^I$ is always tangent to the inflationary actual trajectory, for instance the red trajectory of Fig.\ref{fig:trajectory} after a sharp turn).

Starting from the background equation of motion (\ref{phi_eom_pb}) in the original basis (or from (\ref{eom_bck_gen}) in an arbitrary orthonormal basis), the projection onto the adiabatic (velocity) direction, whose components are denoted $n^I$ (or  $n^m$),  yields
    \begin{equation}
    \label{bckgd_adiab}
        \ddot{\sigma} +3H\dot{\sigma} +V_{,\sigma} = 0\,, \qquad V_{,\sigma} \equiv n^I V_{,I}\,.
    \end{equation}
Since $n^I$ is a unit vector, its time variation is orthogonal to $n^I$, and therefore proportional  to $s^I$, so that one can write
\beq
\label{entropy_vector}
\dot n^I:=\dot\theta s^I
\eeq
where the coefficient of proportionality,  $\dot\theta$, corresponds to  the time derivative of the angle $\theta$ of $n^I$ with respect to the initial field basis.

One can also derive a second order equation of motion for $\theta$, by using the projection
 of the background equation (\ref{phi_eom_pb}) along the entropic direction, which reads
 \beq
 \dot\sigma\dot\theta+V_{,s}=0, \qquad V_{,s} \equiv s^I V_{,I}\,.
 \eeq
 The time derivative of this equation yields
 \beq
 \label{eom_adiab_bckgd}
 \dot\sigma\ddot\theta+\ddot\sigma\dot\theta+\dot\sigma V_{\sigma s}-\dot\theta V_{,\sigma}=0\,,\qquad V_{,\sigma s} \equiv  n^I s^J V_{,IJ}\,,
 \eeq
 where we have used $\dot s^I=-\dot\theta n^I$.
Dividing by $\dot\sigma$ and  using the adiabatic equation (\ref{bckgd_adiab}) to eliminate $V_{,\sigma}$, we finally obtain
\begin{equation}{\label{theta_2nd_eom_gen}}
        \ddot{\theta}+\left(3H+2\frac{\ddot{\sigma}}{\dot{\sigma}}\right)\dot{\theta}+V_{,\sigma s}=0\,.
    \end{equation}
    
As for the perturbations,     their  equations of motion  in the kinematic basis can be obtained directly from (\ref{eom_full}), with (\ref{Z_matrix}). In Fourier space, they read
\begin{eqnarray}
u_{\sigma}''+k^{2}u_{\sigma}-\frac{z''}{z}u_{\sigma} & = & 2\theta'u_{s}'+2\frac{\left(z\theta'\right)'}{z}u_{s},\label{u_sigma_eom_2f}\\
u_{s}''+k^{2}u_{s}+a^{2}\left(V_{,ss} -\dot{\theta}^{2} -H^{2}\left(2-\epsilon\right)\right)u_{s} & = & -2\theta'u_{\sigma}'+2\frac{z'}{z}\theta'u_{\sigma},\label{u_s_eom_2f}
\end{eqnarray}
where $z\equiv a\sqrt{2\epsilon}$. During a sharp turn, the trajectory deviates from the bottom of the valley, and then rapidly oscillates about the curve that follows the bottom of the valley. As a consequence, the angle $\theta$ undergoes strong oscillations  after the turn, as illustrated in Fig.\ref{fig:angles}, which makes the analysis of the above system of equations somewhat complicated.
    
 \begin{figure}[h]
\begin{minipage}{0.45\textwidth}
    \centering
        \includegraphics[width=\textwidth]{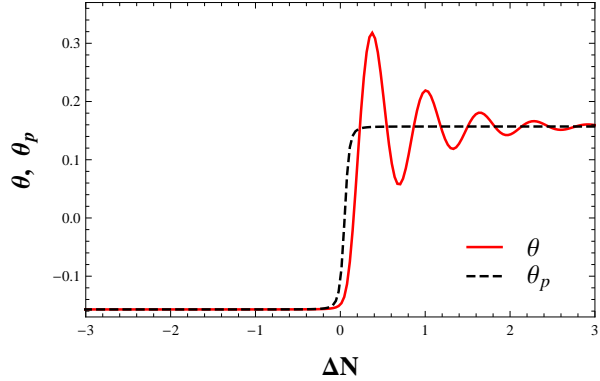}
            \caption{
Behaviour of the kinematic basis (angle $\theta$) and of the potential basis (angle $\theta_m$) during and after a sharp turn.}
            \label{fig:angles}
\end{minipage}
\end{figure}

\subsubsection{Potential/mass basis}
When there exists a mass hierarchy between the various directions in field space, as assumed in the present work, it is also convenient  to use  the ``potential'' basis, i.e. the eigenvectors of the Hessian matrix of the potential. 
 In the potential basis, we  have by construction
     \begin{equation}{\label{V_mn_diag}}
        V_{,mn} = \mathrm{diag}\{\m_{l}^2, \m_{h}^2\},\qquad\text{with } \quad \m_l\ll H\ll \m_h.
    \end{equation}
Let us denote $\theta_p$ the angle between the potential basis and the original field basis.

Now, from the point of view of the potential basis, the 
adiabatic and entropic unit vectors, defined in (\ref{adiab_vector}) and (\ref{entropy_vector}) respectively, can be written in terms of an angle $\psi$
     \begin{equation}{\label{psi_def}}
       n^m = \{\cos\psi,\sin\psi\}\,, \qquad s^m =\{-\sin\psi,\cos\psi\}\,.
    \end{equation}
where  $\psi$  corresponds to the angle of the background velocity with respect to the ``potential basis", i.e.
 \beq
 \theta = \psi +\theta_p\,.
 \eeq
 This implies in particular
    \begin{equation}
   V_{,\sigma\sigma}=\cos^2\psi\, \m_l^2+\sin^2\psi\,  \m_h^2\,, \qquad     V_{,\sigma s} = \frac{1}{2} \left(\m_h^2-\m_l^2 \right)\sin(2\psi)\,.
    \end{equation}

 Substituting this decomposition into (\ref{theta_2nd_eom_gen}), one obtains \cite{GLM}
    \begin{equation}{\label{theta_2nd_eom_2f}}
        \ddot{\psi}+ 3H\left(1+\frac{2\ddot{\sigma}}{3H\dot{\sigma}}\right)\dot{\psi}+ \frac{1}{2} \left(\m_h^2-\m_l^2 \right)\sin(2\psi)= - \ddot{\theta}_p - 3H\left(1+\frac{2\ddot{\sigma}}{3H\dot{\sigma}}\right)\dot{\theta}_p\,,
    \end{equation}
    i.e. an equation of motion for $\psi$, where the source term depends on the angle $\theta_p$.
    The interest of this equation is that, in general, one expects the behaviour of $\theta_p$ to be rather smooth whereas the angle  $\theta$ (and therefore $\psi$) oscillates wildly after a sharp turn, as illustrated in Fig.\ref{fig:angles}. With the above equation, one can thus simply view the evolution of $\psi$ as a response to the change of the potential.

Going back to (\ref{S_canon}), one sees that a slightly different basis is obtained by choosing the (normalized) eigenvectors of the mass matrix (\ref{M_IJ_ori}) rather than of the Hessian of the potential. In principle, the corresponding basis, which we will call the mass basis, differs from the potential basis. However, in the case of  a large mass hierarchy, the potential basis and the mass basis almost coincide  in general because, in the heavy subspace, the first term on the right hand side of (\ref{M_IJ_ori}) is much bigger than the other terms.

Using (\ref{eom_full}), one immediately finds that the equations of motion in the mass basis are given by
    \begin{eqnarray}
u_{l}''+\left(k^{2}+a^{2}m_{l}^{2}-\theta_{m}'^{2}-\frac{a''}{a}\right)u_{l} & = & \theta_{m}''u_{h}+2\theta_{m}'u_{h}',\label{eom_ul_ex}\\
u_{h}''+\left(k^{2}+a^{2}m_{h}^{2}-\theta_{m}'^{2}-\frac{a''}{a}\right)u_{h} & = & -\theta_{m}''u_{l}-2\theta_{m}'u_{l}',\label{eom_uh_ex}
\end{eqnarray}
In contrast with the angle $\theta$ of the kinematic basis, the behaviour of $\theta_m\approx \theta_p$ is very smooth\footnote{Note that $\theta_m$ (and also $\theta_p$) has an oscillatory component, because the trajectory oscillates around the bottom of the valley, but this oscillatory component is in general subdominant with respect to the smooth component.}.

Note that the descriptions in the kinematic basis and in the mass basis are simply related by a rotation of angle $\psi \equiv \theta - \theta_m$, so that, at any time, the adiabatic mode, for instance, is given by
    \begin{equation}
        u_{\sigma}=\cos\psi\,  u_{l}+\sin\psi\,  u_{h}\,,
    \end{equation}
 in terms of the light and heavy modes. Moreover,  long after the turn, the angle $\psi$ relaxes to zero if we assume that there is no subsequent turn. Therefore, the light mode and the adiabatic mode will coincide in the asymptotic limit, which allow us to compute the prediction for the adiabatic power spectrum either by using the adiabatic mode, or the light mode at late time.

\subsection{In-in formalism}
In order to estimate the impact of the turn on the final adiabatic power
spectrum, with respect to the usual featureless quasi scale-invariant
spectrum, we resort to use a perturbative approach. In particular, we
exploit  the in-in formalism, which has been widely used in the recent
literature studying cosmological  perturbations generated during
inflation (\cite{Weinberg:2005vy,Weinberg:2006ac,Seery:2008qj}
and see e.g. \cite{Chen:2010xka}  for a review). In our case, we will need the in-in formalism expressions at first and second order, which can be written respectively as 
\beq
\label{in_in_1}
\langle {\cal O}(\eta)\rangle_{(1)}=2 \Re\left[-i \int_{-\infty}^\eta d\eta'\langle 0| {\cal O}(\eta) H_I (\eta')|0\rangle\right]
\eeq
and
\beq
\label{in_in_2}
\langle {\cal O}(\eta)\rangle_{(2)}=-2 \Re\left[ \int_{-\infty}^\eta d\eta'\int_{-\infty}^{\eta'} d\eta''\langle 0| {\cal O}(\eta) H_I (\eta')H_I (\eta'')|0\rangle\right]+ \int_{-\infty}^\eta d\eta'\int_{-\infty}^{\eta} d\eta''\langle 0|  H_I (\eta') {\cal O}(\eta) H_I (\eta'')|0\rangle\,,
\eeq
where 
 ${\cal O}$ is some operator, and $H_I$ the interaction Hamiltonian.

In the mass basis,  the action  that leads to the  equations of motion (\ref{eom_ul_ex}) and (\ref{eom_uh_ex}) can be obtained directly  from the general expression  (\ref{S_mass}). It can be decomposed as 
	\begin{equation}
		S = \int d\eta\,  d^{3}x \left(  \mathcal{L}^{\mathrm{l}}+ \mathcal{L}^{\mathrm{h}}+ \mathcal{L}^{\theta}\right),
	\end{equation}
with
\begin{eqnarray}
\mathcal{L}^{\mathrm{l,h}} & = & \frac{1}{2}\left[u_{l,h}'^{2}-\left(\partial u_{l,h}\right)^{2}-\left(a^{2}m_{l,h}^{2}-\frac{a''}{a}\right)u_{l,h}^{2}\right],\label{L_s}\\
	\mathcal{L}^{\theta} & = & \frac{1}{2}\theta_{m}'^{2}u_{l}^{2}+\frac{1}{2}\theta_{m}'^{2}u_{h}^{2}+2\theta_{m}'u_{l}u'_{h}+\theta_m''u_{l}u_{h}\,,\label{L_t}
\end{eqnarray}
where the last contribution $\mathcal{L}^{\theta}$ comes from the ${\bf Z}^2$ contribution in  $ \tilde{\bm{M}}$ and from the term $au^{\prime T} {\bf Z} u$ in (\ref{S_mass}), up to  an integration by parts in order to eliminate the time-derivative of $u_l$. This contribution includes all the terms that 
depend explicitly on the variation of the angle $\theta_m$. It contains   a self-coupling term for the light and heavy modes as well as terms  coupling the light and heavy modes. These terms are significant just at the moment of the turn  but are then suppressed very quickly, since the behavior of $\theta_m$ is essentially the same as that of $\theta_p$ (see Fig.~\ref{fig:angles}). 

The  Hamiltonian density associated with the above action is given by
\begin{eqnarray}
\mathcal{H}
&=& \frac{1}{2} \pi_l^2+\frac{1}{2} (\partial u_l)^2 + \frac{1}{2} 
\left(a^2 m^2_l -\frac{a''}{a}  \right)u_l^2
 + \frac{1}{2} \pi_h^2 +\frac{1}{2} (\partial u_h)^2 + \frac{1}{2} 
\left(a^2 m_h^2 -\frac{a''}{a} \right)u_h^2\nonumber\\
&& +\frac32 \theta^{\prime 2}_m  u_l^2-2 \theta_m ' \pi_h u_l -\theta_m '' u_l u_h -\frac12 \theta^{\prime 2}_m u_h^2\,.
\end{eqnarray}
where $\pi_l$ and $\pi_h$ denote the conjugate momenta associated with $u_l$ and $u_h$, respectively.
We now split the Hamiltonian into a free part and an interaction part, given respectively by 
\begin{eqnarray}
\mathcal{H}_0 = \mathcal{H}^{(l)}_0 +\mathcal{H}^{(h)}_0 \equiv \frac{1}{2} \pi_l^2+\frac{1}{2} (\partial u_l)^2 + \frac{1}{2} 
\left(a^2 m^2_l -\frac{a''}{a}  \right)_{(0)} u_l^2
 + \frac{1}{2} \pi_h^2 +\frac{1}{2} (\partial u_h)^2 + \frac{1}{2} 
\left(a^2 m_h^2 -\frac{a''}{a} \right)_{(0)} u_h^2
\end{eqnarray}
and 
\begin{eqnarray}
\label{H_I}
\mathcal{H}_I &=& \mathcal{H}^{(l)} _{I} + \mathcal{H}_{I} ^{(h)} 
+ \mathcal{H}_{I} ^{ (\theta)}\,,
\label{int_hamiltonian_alt}\\
\mathcal{H}^{(l)} _{I}  &=&  \frac12
\Delta \left(a^2  m_l^2 - \frac{a''}{a}  \right)u_l ^2\,,
\label{def_hll_alt}\\
\mathcal{H}_{I} ^{(h)}   &=&   \frac12
\Delta \left(a^2 m_h^2 - \frac{a''}{a}  \right) u_h ^2 \,,\label{def_hhh_alt}\\
\mathcal{H}_{I} ^{ (\theta)} &=& -2\theta_m '  u_l \pi_h   - \theta_m '' u_l u_h+\frac32{\theta_m '}^2 u_l^2-\frac12 {\theta_m '}^2 u_h^2
\,.
\end{eqnarray}
In the perturbative part, we have included not only the terms coming from $\mathcal{L}^{\theta}$ but also the contributions to the effective mass due to the turn (these contributions will be defined more precisely in the next section). 

In the free part, the light and heavy modes are fully decoupled.  Assuming a quasi-de Sitter expansion, i.e. $a=-1/(H\eta)$, the quantum fluctuations are described by 
\begin{eqnarray}
\hat u_{l,h} (\eta, {\bf{k}}) = u_{l,h} (\eta, {\bf{k}})\,  a_{l,h} ({\bf{k}}) 
+ u_{l,h}^* (\eta, -{\bf{k}}) \, a_{l,h} ^\dagger (-{\bf{k}})\,,
\end{eqnarray}
with 
\begin{eqnarray}
&&u_l(\eta, {\bf{k}})  = \frac{e^{-ik\eta}}{\sqrt{2k}}\left(1-\frac{i}{k \eta}\right)\,,
\label{form_vl}\\
&&u_h(\eta, {\bf{k}})  = \frac{\sqrt{\pi}}{2} e^{-\frac{\pi}{2} \nu + i \frac{\pi}{4}}
\sqrt{-\eta} H_{i \nu} ^{(1)} (-k \eta)
\;\;\;\;\;\; {\rm with}\;\;\nu=\sqrt{\frac{m_h^2}{H^2}-\frac{9}{4}}
\label{form_vh}\,.
\end{eqnarray}
where the annihilation and creation operators, $a_{l,h}$ and $a_{l,h} ^\dagger$ respectively, satisfy the usual commutation relations (with the heavy and light sectors  fully independent).

Our final goal will be to compute the  power spectrum $\mathcal{P}_{u_l} $ of the light mode at late times (i.e. in the limit $\eta\to 0$), defined by  \begin{eqnarray}
\langle  0| \hat u_l (0, {\bf{k_1}}) \hat u_l (0, {\bf{k_2}})  | 0\rangle
= (2 \pi)^3 \delta^{(3)} ({\bf{k_1}}+{\bf{k_2}}) \mathcal{P}_{u_l} 
\frac{2 \pi^2}{k_1 ^3}\,.
\end{eqnarray}
Taking into account only the free part of the Hamiltonian, one recovers the standard result 
\beq
\mathcal{P}_{u_l (0)} = \frac{a^2 H^2}{4 \pi^2}\,.
\eeq
In the next two sections, we will compute the corrections to this power spectrum due to the interaction Hamiltonian. As suggested by the decomposition (\ref{H_I}), one can  consider separately  two effects.  The impact of the oscillating background component, due to  $\mathcal{H}^{(l)} _{I} $, will be treated  first\footnote{Note that the contribution $\mathcal{H}_{I} ^{(h)}$ is of no direct relevance for our calculation since we are interested only in the power spectrum of the light mode, which is directly observable, and not that of the heavy mode.}. Finally, we will study the consequences of the couplings induced by the turn, i.e. the term $\mathcal{H}_{I} ^{ (\theta)}$.

\section{Impact of the background oscillations}
\label{section3}
In this section, we first derive an analytical estimate of the background quantity which appears in $\mathcal{H}^{(l)} _{I}$, introduced in 
(\ref{def_hll_alt}). In order to do so, we need to study the evolution of the slow-roll parameter $\epsilon$, which is directly related to the norm $\dot\sigma$  of the velocity vector, since $\epsilon=\dot\sigma^2/(2H^2)$. 
By taking time derivative of the adiabatic background equation (\ref{bckgd_adiab}), 
we get the (exact) equation
\begin{equation}
\frac{\dddot{\sigma}}{\dot{\sigma}}+3H\frac{\ddot{\sigma}}{\dot{\sigma}}+3\dot{H}+V_{,\sigma\sigma}-\dot\theta^2=0\,.
\label{ad_eom_der}
\end{equation}
Using the number of e-folds $N$ instead of cosmic time as the evolution variable, this implies the following \emph{exact} equation of motion
for $\epsilon$:
\begin{equation}
\frac{1}{2\epsilon}\frac{d^{2}\epsilon}{dN^{2}}+\frac{1}{2}\left(3-5\epsilon-\frac{1}{2}\frac{d\ln\epsilon}{dN}\right)\frac{d\ln\epsilon}{dN}-2\epsilon\left(3-\epsilon\right)=S^{(\epsilon)}\,,
\label{epsilon_eom_N}
\end{equation}
 where  
 the source term on the right hand side is defined as
\begin{equation}
S^{(\epsilon)}\equiv \left(\frac{d\theta_{p}}{dN}+\frac{d\psi}{dN}\right)^{2}-\frac{\m_l^2}{H^{2}}-\frac{\m_h^2-\m_l^2}{H^{2}} \sin^2\psi\,.
\label{epsilon_eom_src}
\end{equation}
 The source term depends on the evolution of the angle $\psi$. It  can be determined from the equation (\ref{theta_2nd_eom_2f}), which can be rewritten in terms of e-folding number $N$ as 
\begin{equation}
\frac{d^{2}\psi}{dN^{2}}+3\left(1-\epsilon+\frac{1}{3\epsilon}\frac{d\epsilon}{dN}\right)\frac{d\psi}{dN}+\frac{\m_h^2-\m_l^2}{2H^{2}}\sin\left(2\psi\right)=-\frac{d^{2}\theta_{p}}{dN^{2}}-3\left(1-\epsilon+\frac{1}{3\epsilon}\frac{d\epsilon}{dN}\right)\frac{d\theta_{p}}{dN}.\label{psi_eom_N}
\end{equation}

From now on, we  consider the evolution of $\theta_{p}$ as given since it depends essentially on the shape of the potential, and we will try to estimate the behaviour of $\epsilon$ and of $\psi$ when there is a sharp change of $\theta_p$. A simple way to model this sharp change is to use a Gaussian function of the form
\beq
\frac{d\theta_{p}}{dN}=\Delta\theta\frac{\mu}{\sqrt{2\pi}}e^{-\frac{1}{2}\mu^{2}N^{2}},
\eeq
where 
$\Delta\theta$ is the global ``turning angle" of the trajectory (comparing the trajectory direction well before and after the turn), which we will assume to be small, and
 $\mu$ is a dimensionless quantity  characterizing the ``sharpness" of the turn. 
 As discussed in \cite{GLM}, a  turn is considered to be ``sharp'' when the parameter  $\mu$ is larger than the heavy mass $m_h$. 
In the following, we will restrict ourselves to  the limit where $\mu$ is extremely large, corresponding to an \emph{instantaneous turn} of the trajectory:
\begin{equation}
\label{delta_sharp}
\frac{d\theta_{p}}{dN}=\Delta\theta\, \delta\left(N\right)\,.
\end{equation}
A more general discussion depending on the value of $\mu$ can be found in \cite{GLM}. 
The dimensionless quantity $\Delta\theta$ will play the role of the small parameter in our perturbative treatment of the effects due to the turn. 

\subsection{Evolution of  $\psi$}

Let us start with the equation for $\psi$, which is simpler to solve as it does not depend on $\epsilon$ in the slow-roll approximation.
The evolution of $\psi$ has been analyzed in detail in our previous work \cite{GLM}. Here we simply summarize the results. 

Neglecting $\epsilon$ as well as $\epsilon^{-1}d\epsilon/dN$, and assuming $\psi$ remains small, the equation (\ref{psi_eom_N}) can be approximated by
\beq
\label{d2psi}
\frac{d^{2}\psi}{dN^{2}}+3\frac{d\psi}{dN}+\frac{\m_h^2-\m_l^2}{2H^{2}}\sin\left(2\psi\right)=-\frac{d^{2}\theta_{p}}{dN^{2}}-3\frac{d\theta_{p}}{dN}.
\eeq
Although the Hubble parameter $H$ contains a time-dependent contribution, as follows from the equation (\ref{epsilon_eom_N}) for $\epsilon=(1/H)\dot{}\,$,  we assume that its leading component is constant  during the turning process and $H$ in (\ref{d2psi}) is thus treated as a constant.

By introducing the retarded Green's function associated to the homogeneous part of Eq. (\ref{d2psi}), namely
\begin{equation}
G_{\psi}\left(N,N'\right)=\frac{\sin\left(\hat{\omega}\left(N-N'\right)\right)}{\hat{\omega}}e^{-\frac{3}{2}\left(N-N'\right)}\, \Theta(N-N'),\qquad\hat{\omega}\equiv\sqrt{\frac{m_{h}^{2}}{H^{2}}-\frac{9}{4}}\approx \frac{m_h}{H} \gg1,\label{psi_green_fun}
\end{equation}
it is straightforward to determine the evolution for $\psi$:
\begin{eqnarray}
\psi\left(N\right) & = & -\int^{N}dN'G_{\psi}\left(N,N'\right)\left(\frac{d^{2}\theta_{p}}{dN^{2}}\left(N'\right)+3\frac{d\theta_{p}}{dN}\left(N'\right)\right)\nonumber \\
 & \equiv & -\Delta\theta\,  e^{-\frac{3}{2}N}\, \frac{\cos\left(\hat{\omega}N-\alpha\right)}{\cos\alpha}\, \Theta\left(N\right),\label{psi_sol_der}
\end{eqnarray}
with
\beq
\alpha=\arctan\frac{3}{2\hat{\omega}}\ll 1\,.\label{alpha_def}
\eeq
In the rest of this paper, we will write down our results directly in  the limit $\alpha=0$, for simplicity.
We emphasize that the  approximate solution (\ref{psi_sol_der}) is valid only if $|\Delta\theta |\ll 1$, i.e. the turning angle is small.

\subsection{Evolution of $\epsilon$}
Let us now study the equation of motion (\ref{epsilon_eom_N})  for $\epsilon$. After the turn, the source term (\ref{epsilon_eom_src})  of the equation for  $\epsilon$ is  oscillating as a consequence of  the oscillatory behavior of $\psi$ (\ref{psi_sol_der}) and   $\epsilon$ will thus contain an oscillating component. Although the evolution equation  (\ref{epsilon_eom_N}) is nonlinear with respect to $\epsilon$, in constrast to the equation for $\psi$, one can nevertheless solve it   perturbatively  for small $\Delta\theta$.

Assuming that the oscillatory part of $\epsilon$ is small with respect to its smooth part, i.e. 
\beq
\epsilon\equiv\bar{\epsilon}+\Delta\epsilon,\qquad  \Delta\epsilon\ll \bar{\epsilon}
\eeq
where $\bar{\epsilon}$ is the \emph{smooth part} and $\Delta\epsilon$ the oscillatory part, the linearization of (\ref{epsilon_eom_N}) immediately yields
\begin{equation}
\frac{d^{2}\Delta\epsilon}{dN^{2}}+3\frac{d\Delta\epsilon}{dN}-12\bar{\epsilon}\Delta\epsilon=2\bar{\epsilon}\, S_{\mathrm{osci}}^{(\epsilon)},
\label{epsilon_eom_N_1st_app}
\end{equation}
where the right hand side contains only the oscillatory component of the source term $S^{(\epsilon)}$.
Noting that 
\beq
\frac{d\theta}{dN} = \frac{d\psi}{dN}+\frac{d\theta_{p}}{dN}=\Delta\theta\, \hat{\omega}e^{-\frac{3}{2}N}\sin\left(\hat{\omega}N\right)\,\Theta\left(N\right)\,,
\eeq 
we get
\begin{equation}
S_{\mathrm{osci}}^{(\epsilon)}=-\left(\Delta\theta\right)^{2}\frac{m_{h}^{2}}{H^{2}} e^{-3N}\cos\left(2\hat{\omega}N\right)\,\Theta\left(N\right)\,,
\label{epsilon_src_osci}
\end{equation}
whereas the smooth part is simply $\bar{S}^{(\epsilon)}=-\m_l^2/H^{2}$. 
Again, we treat  $H$ as a  constant. Moreover, in order to solve (\ref{epsilon_eom_N_1st_app}), we will replace $\bar{\epsilon}$ in terms of its constant average value $\epsilon_0$.

We can now solve Eq.~(\ref{epsilon_eom_N_1st_app}) by introducing the 
retarded Green's function 
\[
G_{\epsilon}\left(N,N'\right)=\frac{\sinh\left(w\left(N-N'\right)\right)}{w}e^{-\frac{3}{2}\left(N-N'\right)}\, \Theta\left(N-N'\right),\qquad w=\sqrt{\frac{9}{4}+12\epsilon_{0}}\approx\frac{3}{2}.
\]
The contribution $\Delta\epsilon$ is thus given by 
\begin{eqnarray}
\Delta\epsilon & = & -2\epsilon_{0}\left(\Delta\theta\right)^{2}\frac{m_{h}^{2}}{H^{2}}\, \Theta\left(N\right)\, \int_{0}^{N}dN'\frac{\sinh\left(w\left(N-N'\right)\right)}{w}e^{-\frac{3}{2}\left(N-N'\right)}e^{-3N'}\cos\left(2\hat{\omega}N'\right)\nonumber\\
& = & \frac{1}{2}\epsilon_{0}\left(\Delta\theta\right)^{2}\Theta\left(N\right)e^{-3N}\cos\left(2\hat{\omega}N\right)
+\text{non-osci}\,,
\label{epsilon_osci_fin}
\end{eqnarray}
where we have used the approximation $w\simeq 3/2$ to obtain the last line. 
In (\ref{epsilon_osci_fin}), ``non-osci" denotes an additional non-oscillatory contribution to $\Delta\epsilon$ which is not relevant for our purpose. From the above derivation, the approximate solution for $\Delta\epsilon$ is valid only if $|\Delta\theta|$ is small.

\subsection{Corrections to the power spectrum}

We are now ready to evaluate the corrections to the light power spectrum  due to the resonance between the background oscillations and the mode functions. Neglecting the light mass, the relevant component of the interaction Hamiltonian, introduced in  (\ref{def_hll_alt}),  reduces to the expression (the subscript ``osc" denotes the oscillatory part)
\beq
\mathcal{H}^{(l)} _{I}  =  -\frac{1}{2}\left(\frac{a''}{a}\right)_{\mathrm{osc}}\, u_{l}^{2}\simeq \frac12 a^{2}{H}^{2}\Delta\epsilon_{\mathrm{osc}}\, u_{l}^{2},
\eeq 
where we have used the relation $a''/a={a}^{2}{H}^{2}(2-\epsilon)$. Note that, 
even if  the oscillatory component is negligible in the evolution of $a$ and $H$, it can  become significant in the second derivative of $a$ which appears in the Hamiltonian. 

The oscillatory component of $\Delta\epsilon$, given in (\ref{epsilon_osci_fin}) in terms of the number of e-folds,  can be reexpressed as a function  of the conformal time, using $N\approx H(t-t_*)=-\ln({\eta}/{\eta_{\ast}})$:
\begin{equation}
\Delta\epsilon_{\mathrm{osc}}\left(\eta\right)=\frac{1}{2}\epsilon_{0}\left(\Delta\theta\right)^{2}\, \left(\frac{\eta}{\eta_{\ast}}\right)^{3}\cos\left[2\frac{m_{h}}{H}\ln\left(\frac{\eta}{\eta_{\ast}}\right)\right]\Theta\left[\ln\left(\frac{\eta_{\ast}}{\eta}\right)\right],
\label{epsilon_osci_eta}
\end{equation}
where $t_{\ast}$ and $\eta_{\ast}$ correspond  to the time of the turn.

Substituting  the interaction Hamiltonian
\beq
\label{H_I_osc}
 H_{I}=\frac{1}{2}{a}^{2}{H}^{2}\Delta\epsilon_{\mathrm{osc}} \int d^{3}x\, u_{l}^{2}=\frac{1}{2\eta^{2}}\Delta\epsilon\left(\eta\right)\int\frac{d^{3}p}{\left(2\pi\right)^{3}}u_{l}\left(\eta,\bm{p}\right)u_{l}\left(\eta,-\bm{p}\right)
\eeq
into  the first order expression (\ref{in_in_1}) from the in-in formalism, we get,  
after some manipulations,  the following contribution to the two-point correlation function of the light mode:
\begin{equation}
\Delta \left\langle \hat{u}_{l}\left(\eta,\bm{k}\right)\hat{u}_{l}\left(\eta,\bm{k}'\right)\right\rangle = i\left(2\pi\right)^{3}\delta\left(\bm{k}+\bm{k}'\right)2u_{l}^{\ast2}\left(\eta,k\right)\int_{-\infty}^{\eta}d\eta'\frac{1}{2\eta'^{2}}\Delta\epsilon\left(\eta'\right)u_{l}^{2}\left(\eta',k\right)+\mathrm{c.c.}.\label{inin}
\end{equation}
Similar contributions due to the resonance between the background oscillations and the perturbations have been 
first estimated  in \cite{Chen:2008wn}.
Inserting the expression (\ref{H_I_osc}) into the above formula, one obtains the correction to the light power spectrum, which, in  the late time limit ($\eta\rightarrow0$),  takes the form 
\begin{equation}
\frac{\Delta P\left(k\right)}{P_0\left(k\right)}=-\frac{i}{4}\epsilon_{0}\left(\Delta\theta\right)^{2}\frac{1}{x_{\ast}^{3}}I\left(k\right)+\mathrm{c.c.}\label{Delta_p_int}
\end{equation}
where $P_0$ is the standard single-field power spectrum. We have introduced the quantity $x_{\ast}\equiv-k\eta_{\ast}=k/(a_{\ast}H)$, which expresses the ratio between an arbitrary wavenumber $k$ and the wavenumber $k_\ast\equiv a_\ast H$ that crosses the Hubbe radius precisely at the turn,  as well as  the integral
\begin{equation}
I\left(k\right)=\int_{0}^{x_{\ast}}dx\cos\left[2\frac{m_{h}}{H}\ln\left(\frac{x}{x_{\ast}}\right)\right]e^{2ix}x\left(1+\frac{i}{x}\right)^{2}\,.  \label{Ik}
\end{equation}
In the regime   $m_h\gg H$, the integral $I\left(k\right)$ contains fast-oscillating functions  and  can be estimated by using the ``stationary
phase'' approximation \footnote{The relevant formula is well-known:
$\int dx\, e^{i\phi(x)}\approx\sqrt{\frac{2\pi}{-i\phi''(x_s)}}e^{i\phi(x_s)}$, where $x_s$ satisfies $\phi'(x_s)=0$.
}. 
Writing (\ref{Ik}) as
\begin{equation}\label{Ik_sp}
I\left(k\right) = \frac{1}{2}\int_{0}^{x_{\ast}}dx\, e^{i\phi_{+}\left(x\right)}x\left(1+\frac{i}{x}\right)^{2}+\frac{1}{2}\int_{0}^{x_{\ast}}dx\, e^{i\phi_{-}\left(x\right)}x\left(1+\frac{i}{x}\right)^{2},
\end{equation}
with
\beq
\phi_{\pm}\left(x\right)  \equiv 2x  \pm2\frac{m_{h}}{H}\ln\left(\frac{x}{x_{\ast}}\right)\,.
\label{phase}
\eeq
Since $x>0$, only $\phi_-$ can reach its stationary value for $x_-={m_{h}}/{H}$, provided $x_-$ belongs to  the interval $\left[0,x_{\ast}\right]$, i.e. 
\begin{equation}
k > a_{\ast}m_h,.\label{resonant_cond}
\end{equation}
This implies  that only the modes with sufficiently short wavelengths are affected by the oscillating background and  thus sensitive to the induced  resonant effect.
In this case,
\begin{eqnarray}
I\left(k\right) 
 & \approx & \frac{1}{2}x_{-}\left(1+\frac{i}{x_{-}}\right)^{2}\sqrt{\frac{2\pi}{-i\phi_{-}''\left(x_-\right)}}e^{i\phi_{-}\left(x_{-}\right)}\, \Theta\left(\frac{k}{a_{\ast}m_{h}}-1\right)\,,
 \label{Ik_sol}
\end{eqnarray}
with $\phi_{-}''\left(x_{-}\right)=2{H}/{m_{h}}$.  We thus find that  the correction to the power spectrum (\ref{Delta_p_int}) due to the oscillating background is given by
\begin{equation}
\left(\frac{\Delta P}{P_0}\right)_{\mathrm{osc}} \approx  \frac{\sqrt{\pi}}{4}\epsilon_{0}\left(\Delta\theta\right)^{2}\frac{1}{x_{\ast}^{3}}\left(\frac{m_{h}}{H}\right)^{3/2}\cos\left[2\frac{m_{h}}{H}\ln\left(\frac{k}{a_{\ast}m_{h}}\right)+2\frac{m_{h}}{H}-\frac{\pi}{4}\right]\, \Theta\left(\frac{k}{a_{\ast}m_{h}}-1\right) \,.\label{Delta_P_fin}
\end{equation}
This is the main result of this work. The above expression  means that, just after a the sharp turn, the oscillating background leaves an imprint on the power spectrum with  features that are periodic in $\ln k$, with a frequency $2 m_h/H\gg 1$. These oscillatory features manifest themselves only on  small lengthscales, i.e. for modes large  wavenumber $k>a_{\ast}m_h$. However, the amplitude of these features quickly decreases 
with $k$, because of the factor $1/x_{\ast}^3$. Finally, the global amplitude is also suppressed by the slow-roll parameter $\epsilon_0$ (the turning angle $\Delta \theta$ is also small in our perturbative approach, but this assumption can be relaxed in principle). 
Note that our  result (\ref{Delta_P_fin}) possesses the  same dependence on $k$  as  that obtained  in \cite{Chen:2011zf}, but  the amplitude is smaller in our case. Typically, we find that the effect is proportional to $ (m_h/H)^{3/2}\epsilon$, in contrast with the $(m_h/H)^{7/2}\epsilon$ amplitude obtained in   \cite{Chen:2011zf}. As we explain in the appendix, the result of \cite{Chen:2011zf} relies on a single-field approximation which leads to a calculation that differs from ours.

\section{Impact of the coupling terms arising from the turn}
\label{section4}
We now consider the influence on the power spectrum of the coupling terms that arise directly from the turn, i.e. the terms included in the Hamiltonian  contribution $\mathcal{H}_{I} ^{ (\theta)}$, more specifically 
\beq
\label{H_theta}
\mathcal{H}_{I} ^{ (\theta,1)} =\frac32\theta^{\prime 2}_m u_l^2\,, \qquad \mathcal{H}_{I} ^{ (\theta,2)} = -2\theta_m '  u_l \pi_h   - \theta_m '' u_l u_h
\,.
\eeq
There is also a self-coupling term for the heavy mode in $\mathcal{H}_{I} ^{ (\theta)}$, but since we are interested only in the  power spectrum for the light mode, this term is irrelevant and can be ignored  in our calculation. 

The impact of  these contributions has already been analyzed in details in \cite{GLM}, but by solving directly the equations of motion perturbatively. Here, we wish to  compute the corrections induced on the power spectrum by using the in-in formalism, which has been widely used in the last few years to study cosmological perturbations from inflationary models. We  then show that the results from the in-in formalism are exactly the same as those obtained in \cite{GLM}.

\subsection{Calculation in the in-in formalism}	

From (\ref{H_theta}), the leading order contributions to the two-point
function of the light mode $u_{l}$ due to the direct couplings at the turn arise from the  interaction Hamiltonian
\begin{equation}
H_I^{c}=H_I^{c(1)}+H_I^{c(2)},\label{Hc_gen}
\end{equation}
with (in Fourier space)
\begin{eqnarray}
H_I^{c(1)} & = & \frac{3}{2}\theta_{m}'^{2}\left(\eta\right)\int\frac{d^{3}p}{\left(2\pi\right)^{3}}u_{l}\left(\eta,\bm{p}\right)u_{l}\left(\eta,-\bm{p}\right),\label{Hc1}\\
H_I^{c(2)} & = & -\int\frac{d^{3}p}{\left(2\pi\right)^{3}}u_{l}\left(\eta,\bm{p}\right)\left(\theta_{m}''\left(\eta\right)+2\theta_{m}'\left(\eta\right)\partial_{\eta}\right)u_{h}\left(\eta,-\bm{p}\right),\label{Hc2}
\end{eqnarray}
where $H^{c(1)}$ corresponds to the self-coupling of the light mode $u_{l}$
and $H^{c(2)}$ to the coupling with the heavy mode $u_{h}$. 
It is important to notice that $H^{c(1)}$ is proportional to  $\Delta\theta^2$, whereas $H^{c(2)}$ depends linearly on $\Delta\theta$. However, the two-point correlation function for the light mode must involve at least two $H^{c(2)}$ vertices, and only one $H^{c(1)}$ vertex,  as depicted in  Fig.~\ref{fig:feyman}. 
As a consequence, the two contributions to the two-point function of the light mode will be of the same order, namely quadratic, in the small parameter $\Delta\theta$. Note that the subscripts ``$(1)$'' and ``$(2)$'' thus correspond to the number of vertices in the associated  Feynman-type
diagrams.

\begin{figure}[h]
\begin{minipage}{0.9\textwidth}
    \centering
    \begin{minipage}{0.4\textwidth}
    \centering
        \includegraphics[width=\textwidth]{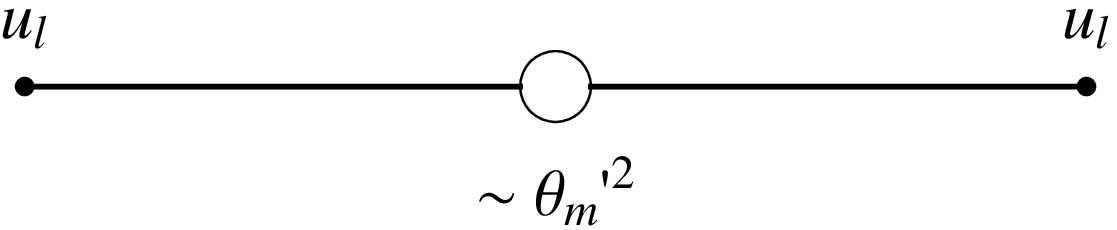}
    \end{minipage}
    ~~~~~~~~~~~~~~~~~~~~~~~~~
    \begin{minipage}{0.4\textwidth}
        \centering
            \includegraphics[width=\textwidth]{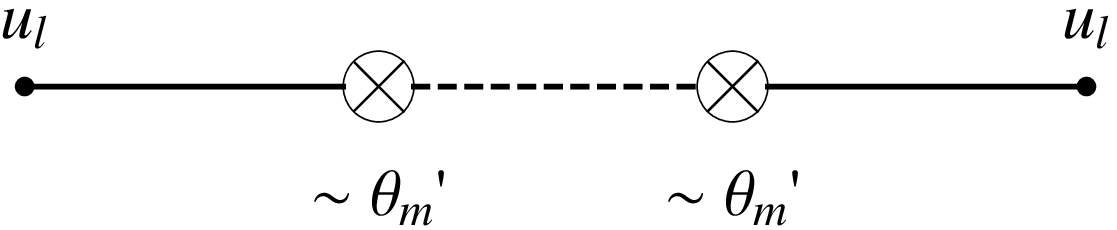}
        \end{minipage}
        \caption{Feynman-type graphs for the leading order contributions due to the turning trajectory to the power spectrum of the light modes. Left panel: contribution from the self-coupling of the light mode itself, which is of order $\sim \theta_m'^2$. Right panel: contribution from the coupling with the heavy mode, which is of order $\sim \theta_m'$. Since the later involves two vertices, both contributions are of the same order of magnitude.}
  \label{fig:feyman}
\end{minipage}
\end{figure}

To compute the first contribution (left in Fig.~\ref{fig:feyman}), one  simply needs to apply the first order expression (\ref{in_in_1}), which yields
\begin{equation}
\left\langle \hat{u}_{l}\left(\eta,\bm{k}_{1}\right)\hat{u}_{l}\left(\eta,\bm{k}_{2}\right)\right\rangle _{(1)}=-2\Re\left[i\int_{-\infty}^{\eta}d\eta'\left\langle \hat{u}_{l}\left(\eta,\bm{k}_{1}\right)\hat{u}_{l}\left(\eta,\bm{k}_{2}\right)\hat{H}^{c(1)}\left(\eta'\right)\right\rangle \right]\label{corr1}
\end{equation}
which is already proportional to $\Delta\theta^2$. By contrast, for  the second contribution (right in Fig.~\ref{fig:feyman}), due to the coupling of $u_l$ with $u_{h}$, one needs the second order expression (\ref{in_in_2}). This gives 
\begin{eqnarray}
\left\langle \hat{u}_{l}\left(\eta,\bm{k}_{1}\right)\hat{u}_{l}\left(\eta,\bm{k}_{2}\right)\right\rangle _{(2)} & = & -2\Re\left[\int_{-\infty}^{\eta}d\eta_{1}\int_{-\infty}^{\eta_{1}}d\eta_{2}\left\langle \hat{u}_{l}\left(\eta,\bm{k}_{1}\right)\hat{u}_{l}\left(\eta,\bm{k}_{2}\right)\hat{H}^{c(2)}\left(\eta_{1}\right)\hat{H}^{c(2)}\left(\eta_{2}\right)\right\rangle \right]\nonumber \\
 &  & +\int_{-\infty}^{\eta}d\eta_{1}\int_{-\infty}^{\eta}d\eta_{2}\left\langle \hat{H}^{c(2)}\left(\eta_{1}\right)\hat{u}_{l}\left(\eta,\bm{k}_{1}\right)\hat{u}_{l}\left(\eta,\bm{k}_{2}\right)\hat{H}^{c(2)}\left(\eta_{2}\right)\right\rangle ,\label{corr2}
\end{eqnarray}
which is proportional to $\Delta\theta^2$ and thus of the same order as (\ref{corr1}). After some straightforward manipulations,
we find that  the correction
to the light power spectrum due to the direct couplings at the turn,  is given, in the late time limit $\eta\rightarrow0$, by the expression
\begin{eqnarray}
\left(\frac{\Delta P}{P_{0}}\right)_{c}\left(k\right) & = & -3i\int_{-\infty}^{0}d\eta\, \theta_m'^{2}\left(\eta\right)\left(u_{l}^{2}\left(\eta\right)-u_{l}^{\ast2}\left(\eta\right)\right)+2\left|\int_{-\infty}^{0}d\eta\,  u_{l}\left(\eta\right)\Gamma\left(\eta\right)\right|^{2}\nonumber \\
 &  & +4\Re\int_{-\infty}^{0}d\eta_{1}\int_{-\infty}^{\eta_{1}}d\eta_{2}\, u_{l}^{\ast}\left(\eta_{1}\right)u_{l}^{\ast}\left(\eta_{2}\right)\Gamma\left(\eta_{1}\right)\Gamma^{\ast}\left(\eta_{2}\right),\label{Delta_P_I}
\end{eqnarray}
where we have defined (the $k$- dependence of the mode functions is not written explicitly, for simplicity)
\[
\Gamma\left(\eta\right):=\theta_m''\left(\eta\right)u_{h}\left(\eta\right)+2\theta_m'\left(\eta\right)u_{h}'\left(\eta\right).
\]
We refer the reader to our previous paper \cite{GLM} to find quantitative estimates of the above expression, since it is equivalent to our previous result as we show next.

\subsection{Comparison with the Green's function approach}
In our previous work \cite{GLM}, we have obtained the correction to the power spectrum by working directly with the equations of motion for the perturbations (\ref{eom_ul_ex})-(\ref{eom_uh_ex}). For example, let us write the equation of motion (\ref{eom_ul_ex}) for the light mode (with $m_l=0$) as
\beq
u_l''+\left(k^2-\frac{a''}{a}\right)=S_l+S_h\,,
\eeq
with the two source terms
\beq
S_l=\theta_m^{\prime 2}u_l,\qquad S_h=\theta_m''u_h+2\theta_m'u_h'\,.
\eeq
Note that  $S_l$ is linear in $\Delta\theta$ whereas $S_h$ is  quadratic, similarly to  $H^{c(2)}$ and $H^{c(1)}$ introduced above.
Treating the source terms perturbatively, one can write the solution of the above equation in the form\beq
u_l\simeq \lambda_l \, u_{l(0)}+C u_{l(0)}+ D u_{l(0)}^*\,,
\eeq
where $\lambda_l$ is some coefficient depending on the initial conditions\footnote{In principle, the homogeneous solution contains also a term $u_{l(0)}^*$ but we  only consider initial conditions (Bunch-Davies vacuum) where this component vanishes.}
and the perturbative coefficients are given by
\beq
\label{C_D}
C=i \int u_{l(0)}^* \left(S_l+S_h\right), \qquad D=-i \int u_{l(0)} \left(S_l+S_h\right)\,.
\eeq
Similarly,  the perturbative solution of the equation of motion (\ref{eom_uh_ex}) for  the heavy mode is 
\beq
u_h\simeq \lambda_h \, u_{h(0)}+E u_{h(0)}+ F u_{h(0)}^*\,,
\eeq
where the  coefficients $E$ and $F$ can we written in a form analogous to (\ref{C_D}). 

In order to compute the power spectrum, one must   take into account the fact that  the initial Bunch-Davies vacua for the light and heavy modes are uncorrelated and therefore  sum in quadrature the power spectra obtained separately with the initial conditions $(\lambda_l=1, \lambda_h=0)$ and $(\lambda_l=0, \lambda_h=1)$.  The total power spectrum is thus
\beq
\frac{P}{P_0}=\left|1+C_1-D_1\right|^2+\left|C_2-D_2\right|^2
\eeq
where the first term corresponds to the initial conditions $(\lambda_l=1, \lambda_h=0)$  and the second to $(\lambda_l=0, \lambda_h=1)$. Since we have solved the equations of motion perturbatively, we must expand the above expression with respect to the small parameter $\Delta\theta$.
The coefficients $C_1$ and $D_1$ contain two types of components: a term  coming from $S_l$ with $u_l=u_{l(0)}$, and a  term coming from $S_h$ with $u_h=E_1 u_{h(0)}+ F_1 u_{h(0)}^* $, and both terms are of order $\Delta\theta^2$. By contrast,  $C_2$ and $D_2$ start  at order $\Delta\theta$. Consequently, the leading correction of the power spectrum is of order $\Delta\theta^2$ and is given by
\beq
\frac{\Delta P}{P_0}=2 \Re(C_1-D_1)+\left|C_2-D_2\right|^2\,.
\eeq
The explicit calculation gives \cite{GLM}
\begin{eqnarray}
\left(\frac{\Delta P}{P_0}\right)_\theta\left(k\right) 
 & = & -3i\int_{-\infty}^{0}d\eta\, \theta_m'^{2}\left(\eta\right)\left(u_{l}^{2}\left(\eta\right)-u_{l}^{\ast2}\left(\eta\right)\right)+4\left|\int_{-\infty}^{0}d\eta\, \Re u_{l}\left(\eta\right)\Gamma\left(\eta\right)\right|^{2} \nonumber\\
 &  & -\int_{-\infty}^{0}d\eta_{1}\int_{-\infty}^{\eta_{1}}d\eta_{2}\left(u_{l}\left(\eta_{1}\right)+u_{l}^{\ast}\left(\eta_{1}\right)\right)\left(u_{l}\left(\eta_{2}\right)-u_{l}^{\ast}\left(\eta_{2}\right)\right)\left[\Gamma\left(\eta_{1}\right)\Gamma^{\ast}\left(\eta_{2}\right)-\Gamma^{\ast}\left(\eta_{1}\right)\Gamma\left(\eta_{2}\right)\right].
\end{eqnarray}
The above form can be obtained  by performing integrations by parts and regrouping the terms in the relevant expressions (Eq.(5.16)-Eq.(5.19)) in \cite{GLM}.

The difference between this result and the expression obtained via the in-in formalism (\ref{Delta_P_I}) can be written as 
	$I_{1}+2\Re I_{2}$,
with
	\begin{eqnarray}
	I_{1} & := & -\int_{-\infty}^{0}d\eta_{1}\int_{-\infty}^{0}d\eta_{2}\left[u_{l}\left(\eta_{1}\right)u_{l}^{\ast}\left(\eta_{2}\right)-u_{l}^{\ast}\left(\eta_{1}\right)u_{l}\left(\eta_{2}\right)\right]\Gamma\left(\eta_{1}\right)\Gamma^{\ast}\left(\eta_{2}\right)\nonumber \\
	 &  & +\int_{-\infty}^{0}d\eta_{1}\int_{-\infty}^{\eta_{1}}d\eta_{2}\left[u_{l}\left(\eta_{1}\right)u_{l}^{\ast}\left(\eta_{2}\right)-u_{l}^{\ast}\left(\eta_{1}\right)u_{l}\left(\eta_{2}\right)\right]\left[\Gamma\left(\eta_{1}\right)\Gamma^{\ast}\left(\eta_{2}\right)-\Gamma^{\ast}\left(\eta_{1}\right)\Gamma\left(\eta_{2}\right)\right],\label{I_1}\\
	I_{2} & := & \int_{-\infty}^{0}d\eta_{1}\int_{-\infty}^{0}d\eta_{2}u_{l}^{\ast}\left(\eta_{1}\right)u_{l}^{\ast}\left(\eta_{2}\right)\Gamma\left(\eta_{1}\right)\Gamma^{\ast}\left(\eta_{2}\right)\nonumber \\
	 &  & -\int_{-\infty}^{0}d\eta_{1}\int_{-\infty}^{\eta_{1}}d\eta_{2}u_{l}^{\ast}\left(\eta_{1}\right)u_{l}^{\ast}\left(\eta_{2}\right)\left[\Gamma\left(\eta_{1}\right)\Gamma^{\ast}\left(\eta_{2}\right)+\Gamma^{\ast}\left(\eta_{1}\right)\Gamma\left(\eta_{2}\right)\right].\label{I_2}
	\end{eqnarray}
Using the trick
	\begin{equation}
	\int_{-\infty}^{0}d\eta_{1}\int_{-\infty}^{0}d\eta_{2}\,\Phi\left(\eta_{1},\eta_{2}\right)=\int_{-\infty}^{0}d\eta_{1}\int_{-\infty}^{\eta_{1}}d\eta_{2}\,\Phi\left(\eta_{1},\eta_{2}\right)+\int_{-\infty}^{0}d\eta_{1}\int_{-\infty}^{\eta_{1}}d\eta_{2}\,\Phi\left(\eta_{2},\eta_{1}\right),
	\end{equation}
to rewrite the first lines in (\ref{I_1}) and (\ref{I_2}) respectively, one can show immediately that $I_1=I_2=0$, which implies that the Green's function approach and the in-in formalism are exactly equivalent, as could have been expected (see e.g. \cite{Weinberg:2008mc}).

\section{Conclusions}
In this work, we have considered in detail the impact of a sudden turn of the inflationary trajectory in a two-dimensional potential that contains a light direction and a heavy direction. In contrast to many treatments in the literature, our approach goes beyond a description in terms of  a single effective degree of freedom and takes into account both the light and heavy modes. Moreover, we have tried to rely as much as possible on an analytical treatment: even if it requires some approximations, it gives more physical insight into the phenomena than a numerical resolution. 

We have considered separately two effects. One is the impact of the direct coupling between the light and massive modes at the instant of the turn. Although we had already treated this aspect in our previous work \cite{GLM}, we have computed this effect here by using a different route, namely  the in-in formalism  that has become ubiquitous in the recent literature on perturbations. In this way, we have confirmed our previous conclusions, based on the more traditional use of Green's function methods. The second aspect is the resonant effect on the power spectrum of the tiny periodic component in the background due to the oscillations along the massive direction after the turn. In contrast with some previous estimates, our analytical treatment indicates that the amplitude of this effect is in general small, since it is suppressed by the slow-roll parameter $\epsilon$ (and enhanced by a factor $(m_h/H)^{3/2}$). The dependence on the wavenumber is the same as calculated previously.  
We believe that the difference between previous analytical works and ours is due to the single-field approximation adopted in the former. As discussed in more details in our appendix (see also \cite{CGLM}), such truncation to a single degree of freedom depends crucially on the choice of gauge and the choice of variables. 
Although our results indicate that the resonant effect in the simplest models will be very difficult to detect, it is worth mentioning that  significant resonant effects could be generated in models involving derivative couplings between the inflaton and the heavy field~\cite{Saito:2012pd,Kobayashi:2012kc,Saito:2013aqa}.

\acknowledgements
We would like to thank Xingang Chen, whose stimulating questions initiated this project,  for  fruitful and challenging discussions and correspondence. 
We  are also  grateful to  Ryo Saito, who  pointed out the difference between the single-field approximation in the flat gauge
and that in the uniform-light field gauge, for very fruitful discussions that helped the writing of the  Appendix in this paper.
We finally thank Thorsten Battefeld for very instructive discussions.
X.G. and D.L. were partly supported by ANR (Agence Nationale de la Recherche) grant ``STR-COSMO" ANR-09-BLAN-0157-01. S.M. was supported by IN2P3.

\appendix
\section{Comparison of our variables  with previous works}

Let us write the scalarly perturbed metric in the form 
\beq
ds^2=-(1+2\phi)dt^2+2\partial_i B \, dt \, dx^i+ a^2(t)\left[\left(1-2\psi\right)\delta_{ij}+2\partial_i\partial_j E\right]dx^i dx^j\,.
\eeq

\subsection{Single scalar field}
Let us first recall the usual notation for the perturbations in single field inflation, where the scalar field is denoted $\phi$.
Combining the metric perturbation $\psi$ and  the scalar field perturbation $\delta\phi$, one can construct the gauge-invariant quantity
\beq
Q:=\delta\phi+ \frac{\dot\phi}{H} \psi.
\eeq
In the flat gauge ($\psi=0$), the field fluctuation  coincides with  $Q$. Alternatively, one can work with the other gauge-invariant quantity 
\beq
\label{R}
{\cal R}:=\frac{H}{\dot\phi} Q= \psi+ \frac{H}{\dot\phi} \delta\phi
\eeq
In the uniform field gauge (i.e. $\delta\phi=0$), the curvature perturbation $\psi$ coincides with ${\cal R}$. 
Note that the quantity $\R$, up to a sign,  is often denoted $\zeta$ in the literature \cite{Maldacena:2002vr,Chen:2010xka}\footnote{It should however not be confused with $\zeta_{\rm Bardeen} \equiv  -\psi + H{\delta \rho}/{\rho}$. The quantity $\zeta_{\rm Bardeen}$ differs  from $\R$, although they coincide on super-Hubble scales.}. 

The quadratic action governing the dynamics of the scalar degree of freedom, can be written in terms of $Q$,
\beq
S_Q=\frac12\int dt d^3x\, a^3\left\{\dot Q^2-\frac{1}{a^2}\partial_iQ\partial^i Q-\left[V''-\frac{1}{a^3}\frac{d}{dt}\left(\frac{a^3}{H}\dot\phi^2\right)\right]Q^2\right\}
\eeq
or in terms of $\R$,
\beq
\label{S_R}
S_\R=\frac12\int dt d^3x\, a^3 \frac{\dot{\phi}^2}{2H^2}\left\{\dot \R^2-\frac{1}{a^2}\partial_i\R\partial^i \R\right\}
\eeq

\subsection{Two-field models}
Let us now discuss  the perturbations in models with two scalar fields. From the two scalar field perturbations $\delta\phi_I$, one can construct two gauge-invariant quantities 
\beq
\label{Q_I}
Q_I:=\delta\phi_I+ \frac{\dot\phi_I}{H} \psi.
\eeq

One can also define the global gauge-invariant curvature perturbation
\beq
{\cal R}=\frac{H}{\|\dot{\vec\phi}\|^2} \dot\phi_I Q_I= \psi+ \frac{H}{\|\dot{\vec\phi}\|^2} \dot\phi_I \delta\phi_I\,.
\eeq

\subsubsection{Kinematic basis}
Instead of the initial field basis, one can decompose the perturbations in the kinematic basis.

The first vector, which we denote $e^I_\parallel$, is parallel to  the instantaneous velocity along the inflationary trajectory, and the second vector, denoted $e^I_\perp$, is orthogonal to the first. The corresponding gauge-invariant  quantities in this basis are denoted respectively
\beq
 (Q_\parallel, Q_\perp) \qquad{\rm kinematic}\ {\rm basis}
\eeq 
Note that 
\beq
{\cal R}=\frac{H}{\|\dot{\vec\phi}\|} Q_\parallel
\eeq
The quantity 
\beq
{\cal S}= \frac{H}{\|\dot{\vec\phi}\|} Q_\perp
\eeq
corresponds to the (instantaneous) isocurvature perturbation. One can also work directly with  ${\cal R}$ and $Q_\perp$, as in Ref. \cite{Cespedes:2012hu} (where $Q_\perp$ is named ${\cal F}$).

\subsubsection{``potential'' basis}
One can choose for the basis the two eigenvectors of the matrix $\partial_I\partial_J V$. The first, corresponding to the smallest eigenvalue, is the light vector  denoted $e^I_l$; the second represents the heavy direction and is denoted $e^I_h$. The corresponding gauge-invariant  quantities in this basis are
\beq
\label{Q_potential}
(Q_l, Q_h) \qquad{\rm potential}\ {\rm basis}
\eeq

In general, one goes from one basis to another by a simple rotation.
An important point is that, {\it before and long after the turn}, the two above bases coincide. 

\subsubsection{Uniform light-field gauge}
In single field inflation, it is sometimes convenient to use the uniform-field gauge leading, as we saw above, to  the action (\ref{S_R}). One possible extension of this gauge in the two-field context is  the uniform \emph{light-field} gauge ($E=0, \delta\phi_l=0$), which has been used in \cite{Chen:2011zf}.
In this description, the two degrees of freedom are
\beq
\zeta\equiv \psi= \frac{H}{\dot{\phi}_l}Q_l\,, \qquad  \delta\sigma \equiv \delta\phi_h\,.
\eeq
In \cite{Chen:2011zf}, the resonant effect was computed by truncating the quadratic action expressed in terms  of $\zeta$ and $\delta\sigma$. Keeping  the terms that depend only on $\zeta$ (i.e. settting $\delta\sigma=0$) leads to an action where the kinetic and gradient terms  for $\zeta$ are both
 proportial to $\epsilon=(\dot\phi_l^2+\dot\phi_h^2)/(2H^2)$.

However, the uniform-light-field gauge is very special in the sense that the fluctuation $\delta \sigma$ in fact depends on both the gauge-invariant perturbations $Q_h$ and $Q_l$:
\beq
\delta\sigma=Q_h - \frac{\dot{\phi}_h}{\dot{\phi}_l}Q_l\,.
\eeq
In order to disentangle the light and heavy degrees of freedom, it is useful to write the quadratic action for the perturbations 
(\ref{S_canon}) in terms 
of $\zeta$ and of $Q_h$. One finds
\beq
S=\int dt d^3x \, a^3\left\{ \frac{\dot\phi_l^2}{2H^2} \dot\zeta^2- \frac{\dot\phi_l^2}{2H^2} \frac{(\partial \zeta)^2}{a^2}+\frac12 \dot Q_h^2- \frac12\frac{(\partial Q_h)^2}{a^2} +\frac12\mu^2_\zeta \zeta^2+\frac12\mu_{Q_h}^2Q_h^2+\lambda \zeta Q_h\
\right\}\,,
\eeq
with
\begin{eqnarray}
\mu^2_\zeta &\equiv& \frac{(3-\epsilon)}{H^2} \dot{\phi}_l ^2 \dot{\phi}_h ^2
+\frac{\dot{\phi}_l \dot{\phi}_h}{H^3} \left(\dot{\phi}_l V_{,h} + 
\dot{\phi}_h V_{,l}\right), \qquad \mu^2_{Q_h}\equiv 
-\left(V_{,hh} +2 \frac{\dot{\phi}_h V_{,h}}{H}
+(3-\epsilon)\dot{\phi}_h^2\right),\nonumber\\
\lambda &\equiv& 2 \frac{\dot{\phi}_l}{H}\left[
\left(V_{,l} \frac{\dot{\phi}_h}{H}+V_{,h}\frac{\dot{\phi}_l}{H}\right)
+(3-\epsilon) \dot{\phi}_l \dot{\phi}_h\right]\,.
\end{eqnarray}
If one truncates the above action to keep only the light degree of freedom $\zeta$, one sees that the coefficient in front of $\dot\zeta^2$ and $(\partial \zeta)^2$ is  $\dot\phi_l^2/(2H^2)$ instead of $\epsilon$: the $\dot\phi_h^2$ contribution, which is the main oscillating term, is not present in our case.  More details and a broader discussion of this issue will be presented elsewhere~\cite{CGLM}.


\end{document}